\documentstyle[12pt,fleqn]{article}

\def\thebibliography#1{\section*{References}\list
  {[\arabic{enumi}]}{\settowidth\labelwidth{#1}\leftmargin\labelwidth
    \advance\leftmargin\labelsep
    \usecounter{enumi}}
    \def\newblock{\hskip .11em plus .33em minus .07em}
    \sloppy\clubpenalty4000\widowpenalty4000
    \sfcode`\.=1000\relax}

\def\op#1{\mathop{\fam0 #1}\limits}

\newcommand{\id}{{\rm Id\,}}
\newcommand{\Ker}{{\rm Ker\,}}
\newcommand{\nm}[1]{\mid {#1}\mid}

\newcommand{\beq}{\begin{equation}}
\newcommand{\eeq}{\end{equation}}
\newcommand{\ben}{\begin{eqnarray}}
\newcommand{\een}{\end{eqnarray}}
\newcommand{\be}{\begin{eqnarray*}}
\newcommand{\ee}{\end{eqnarray*}}
\newcommand{\bea}{\begin{eqalph}}
\newcommand{\eea}{\end{eqalph}}

\newcommand{\cA}{{\cal A}}

\newcommand{\cP}{{\cal P}}
\newcommand{\cV}{{\cal V}}

\newcommand{\cR}{{\cal R}}
\newcommand{\cL}{{\cal L}}

\newcommand{\cH}{{\cal H}}
\newcommand{\cN}{{\cal N}}
\newcommand{\cF}{{\cal F}}
\newcommand{\cS}{{\cal S}}

\newcommand{\al}{\alpha}

\newcommand{\dl}{\delta}
\newcommand{\la}{\lambda}

\newcommand{\f}{\phi}
\newcommand{\om}{\omega}

\newcommand{\m}{\mu}

\newcommand{\G}{\Gamma}

\newcommand{\si}{\sigma}
\newcommand{\w}{\wedge}

\newcommand{\wh}{\widehat}
\newcommand{\ol}{\overline}
\newcommand{\dr}{\partial}

\newcommand{\ot}{\otimes}

\newcommand{\nw}[1]{[{#1}]}
\newcommand{\der}{\rm Der}
\newcommand{\pr}{{\rm pr}}

\newcounter{eqalph}
\newcounter{equationa}
\newcounter{theorem}
\newcounter{remark}
\newcounter{proposition}
\newcounter{lemma}
\newcounter{corollary}
\newcounter{definition}

\newenvironment{eqalph}{\stepcounter{equation}
\setcounter{equationa}{\value{equation}}
\setcounter{equation}{0}

\begin{eqnarray}}{\end{eqnarray}\setcounter{equation}{\value{equationa}}}

\def\theremark{\arabic{remark}}

\def\thedefinition{\arabic{definition}}

\begin{document}
\hbox{}

\begin{center}
{\large\bf THE KOSZUL--TATE COHOMOLOGY IN

COVARIANT HAMILTONIAN FORMALISM}
\bigskip

{\sc  LUIGI MANGIAROTTI}\footnote{E-mail: mangiaro@camserv.unicam.it}

{\small
 Department of Mathematics and Physics, University of Camerino, 

62032
Camerino (MC), Italy }
\medskip

{\sc GENNADI SARDANASHVILY}\footnote{E-mail address: sard@grav.phys.msu.su}

{\small
 Department of Theoretical Physics, Physics Faculty, Moscow State
University,

 117234 Moscow, Russia}
\bigskip

\end{center}

\begin{small}

We show that, in the framework of covariant Hamiltonian field theory, a
degenerate almost regular quadratic Lagrangian $L$ admits a complete set of
non-degenerate Hamiltonian forms such that solutions of the corresponding
Hamilton equations, which live in the Lagrangian constraint space, exhaust
solutions of the Euler--Lagrange equations for $L$. We obtain the
characteristic splittings of the configuration and momentum phase bundles. Due
to the corresponding projection operators, the Koszul--Tate resolution of the
Lagrangian constraints for a generic almost regular quadratic Lagrangian is
constructed in an explicit form.
\end{small}
\bigskip

\section{Introduction}

The covariant Hamiltonian approach to field theory has been
vigorously developed from the seventies (see [1-3] for a
survey). This is the Hamiltonian counterpart of Lagrangian field theory on
fibre bundles $Y\to X$, which looks promising for quantization [4]. If
dim$X=1$, covariant Hamiltonian formalism provides the adequate description
of Hamiltonian time-dependent mechanics [5,6]. Here, we aim to
apply some elements of the classical BRST technique for Hamiltonian
systems to covariant Hamiltonian theory.

Let us recall that, given a fibre bundle
$Y\to X$ coordinated by $(x^\la,y^i)$,  a first order Lagrangian $L$ of
fields is defined as a horizontal density
\beq
L=\cL(x^\la,y^i,y^i_\la)\om, \quad \om=dx^1\w\cdots dx^n,
\quad n=\dim X,
\label{cmp1}
\eeq
on the jet bundle $J^1Y\to X$ seen as a finite-dimensional configuration
space of fields and coordinated by $(x^\la,y^i,y^i_\la)$. The corresponding
Euler--Lagrange equations take the coordinate form
\beq
(\dr_i- d_\la\dr^\la_i)\cL=0, \qquad d_\la=\dr_\la +y^i_\la\dr_i
+y^i_{\la\m}\dr^\m_i.
\label{b327} 
\eeq
Every Lagrangian $L$ (\ref{cmp1}) yields the Legendre map 
\be
\wh L:J^1Y \op\to_Y \Pi, \qquad p^\la_i\circ\wh L =\pi^\la_i=\dr_i^\la\cL,
\ee
of $J^1Y$
to the Legendre bundle 
\be
\Pi=\op\w^nT^*X\op\ot_YV^*Y\op\ot_YTX, 
\ee
equipped with the holonomic coordinates $(x^\la,y^i,p^\la_i)$ and  
seen as a momentum
phase space of fields [7,8]. 
Hamiltonian dynamics on $\Pi$ is phrased in terms of Hamiltonian forms
\be
 H= p^\la_i dy^i\w \om_\la -\cH(x^\la,y^i,p^\la_i)\om, \qquad
\om_\la=\dr_\la\rfloor\om, 
\ee
and the corresponding covariant Hamilton equations
\beq
y^i_\la= \dr_\la^i\cH, \qquad p^\la_{\la i}= - \dr_i\cH. \label{b4100}
\eeq

In the case of hyperregular Lagrangians, Lagrangian formalism and
covariant Hamiltonian formalism are equivalent. For any hyperregular
Lagrangian $L$, there exists a unique Hamiltonian form $H$ such that the
Euler--Lagrange equations (\ref{b327}) for $L$ are equivalent to the
Hamilton equations (\ref{b4100}) for $H$. The case of degenerate Lagrangians
is more intricate. Let us restrict our consideration to almost
regular Lagrangians $L$, i.e.,
(i) the Lagrangian constraint space $N_L=\wh L(J^1Y)$ is a closed imbedded
subbundle of the Legendre bundle $\Pi\to Y$, (ii) the Legendre map $\wh
L:J^1Y\to N_L$ is a fibred manifold, and (iii) the inverse image $\wh
L^{-1}(q)$ of  any point $q\in N_L$ is a connected submanifold of $J^1Y$.
At least locally, one can assign to $L$ a complete set of Hamiltonian forms $H$
such that there is one-to-one correspondence between solutions of the Hamilton
equations for
$H$ from this set, which live in the Lagrangian constraint space $N_L$, and
solutions of the Euler--Lagrange equations for $L$ (see
\cite{sard95,book,giac99} for a detailed exposition). It is important  that
these Hamiltonian forms $H$ are not necessarily degenerate. 

In the present work, the case of almost regular quadratic
Lagrangians, appropriate for application to many physical models, is studied
in detail. We obtain a complete set of
non-degenerate Hamiltonian forms with the above mentioned properties for a
generic almost regular quadratic Lagrangian. 
We show that, in this case, the
Legendre bundle
$\Pi$ admits the characteristic splitting 
$\Pi=\Ker\si\oplus N_L$ (\ref{N20}).
Using the corresponding projection operators, we construct the Koszul--Tate
resolution for the Lagrangian constraints $N_L$ of a generic almost
regular quadratic Lagrangian $L$ in an explicit form.

\section{Quadratic degenerate systems}

Given a fibre bundle $Y\to X$,
let us consider a  quadratic Lagrangian $L$ which has the coordinate
 expression
\beq
\cL=\frac12 a^{\la\m}_{ij} y^i_\la y^j_\m +
b^\la_i y^i_\la + c, \label{N12}
\eeq
where $a$, $b$ and $c$ are local functions on $Y$.
This property is
coordinate-independent due to the affine transformation law of coordinates
$y^i_\la$. The associated Legendre map 
\beq
p^\la_i\circ\wh L= a^{\la\m}_{ij} y^j_\m +b^\la_i \label{N13}
\eeq
is an affine morphism over $Y$. It defines the corresponding linear
morphism
\beq
\ol L: T^*X\op\otimes_YVY\op\to_Y\Pi,\qquad p^\la_i\circ\ol
L=a^{\la\m}_{ij}\ol y^j_\m, \label{N13'}
\eeq
where $T^*X\ot_Y VY$ is the underlying vector bundle of the affine jet bundle
$J^1Y\to Y$ and 
$\ol y^j_\mu$ are  bundle coordinates on it.

Let the Lagrangian $L$ (\ref{N12}) be almost regular, i.e.,
the matrix function $a^{\la\m}_{ij}$ is of constant rank. Then
the Lagrangian constraint space $N_L$ 
(\ref{N13}) is an affine subbundle of the Legendre bundle $\Pi\to Y$, modelled
over the vector subbundle $\ol N_L$ (\ref{N13'}) of  $\Pi\to
Y$. 
Hence, $N_L\to Y$ has a global section. For the sake of simplicity, let us
assume that it is the canonical
zero section $\wh 0(Y)$ of $\Pi\to Y$. Then $\ol N_L=N_L$.
Accordingly, the kernel
of the Legendre map (\ref{N13})  is an affine
subbundle of the affine jet bundle $J^1Y\to Y$, modelled over the kernel of
the linear morphism $\ol L$ (\ref{N13'}). Then there exists a connection 
\beq
\G: Y\to \Ker\wh L\subset J^1Y, \qquad
a^{\la\m}_{ij}\G^j_\m + b^\la_i =0, \label{N16}
\eeq
on $Y\to X$.
Connections (\ref{N16}) constitute an affine space modelled over the linear
space of soldering forms $\f$ on $Y\to X$ satisfying the conditions
\beq
a^{\la\m}_{ij}\f^j_\m =0, \qquad \f^i_\la b^\la_i=0. \label{cmp21}
\eeq

The key point of our consideration is a linear bundle
map
\beq
\si: \Pi\op\to_Y T^*X\op\otimes_YVY, \qquad
\ol y^i_\la\circ\si =\si^{ij}_{\la\m}p^\m_j, \label{N17}
\eeq
such that $\ol L\circ\si|_{N_L}= \id N_L$.
It is a solution of the algebraic equations
\be
a^{\la\mu}_{ij}\si^{jk}_{\mu\al}a^{\al\nu}_{kb}=a^{\la\nu}_{ib}.
\ee
The matrix $a$ in the Lagrangian $L$ (\ref{N12}) can be seen as a
global section of constant rank of the tensor bundle 
\be
\op\w^n T^*X\op\ot_Y[\op\vee^2(TX\op\ot_Y V^*Y)]\to Y.
\ee
By virtue of the well-known theorem on the splitting of an exact
sequence of vector bundles, there exists the bundle splitting 
\beq
T^*X\op\ot_Y VY=\Ker a\op\oplus_Y E', \label{mm46}
\eeq
together with a (non-holonomic) atlas of this bundle such that transition
functions  of
$\Ker a$ and $E'$ are independent. Since $a$ is a non-degenerate section of
$\op\w^n T^*X\op\ot_Y(\op\vee^2E'^*)\to Y$, there exists an atlas of $E'$ such
that $a$ is brought into a diagonal matrix with non-vanishing
components
$a^{AA}$. Due to the splitting (\ref{mm46}), we have the corresponding
bundle splitting
\be
TX\op\ot_Y V^*Y=(\Ker a)^*\op\oplus_Y E'^*.
\ee
Then the desired map $\si$ is represented by a direct sum $\si_1\oplus\si_0$
of an arbitrary section $\si_1$ of the fibre bundle
\be
\op\w^n TX\op\ot_Y(\op\vee^2\Ker a)\to Y
\ee
and the section
$\si_0$ of the fibre bundle 
\be
\op\w^n TX\op\ot_Y(\op\vee^2E')\to Y
\ee
which has non-vanishing components $\si_{AA}=(a^{AA})^{-1}$ with respect to
the above mentioned atlas of $E'$.  Moreover,
$\si$ satisfies the additional relations
\be
\si_0=\si_0\circ\ol L\circ\si_0, \quad a\circ\si_1=0, \quad \si_1\circ a=0.
\ee

With the map (\ref{N17}), we have the splittings
\bea
&& J^1Y=\cS(J^1Y)\op\oplus_Y \cF(J^1Y)=\Ker\wh L\op\oplus_Y{\rm Im}(\si\circ
\wh L), \label{N18} \\
&& y^i_\la=\cS^i_\la+\cF^i_\la= [y^i_\la
-\si^{ik}_{\la\al} (a^{\al\m}_{kj}y^j_\m + b^\al_k)]+
[\si^{ik}_{\la\al} (a^{\al\m}_{kj}y^j_\m + b^\al_k)], \label{b4122}
\eea
\bea
&& \Pi=\cR(\Pi)\op\oplus_Y\cP(\Pi)=\Ker\si_0 \op\oplus_Y N_L, \label{N20} \\
&& p^\la_i = \cR^\la_i+\cP^\la_i= [p^\la_i -
a^{\la\m}_{ij}\si^{jk}_{\m\al}p^\al_k] +
[a^{\la\m}_{ij}\si^{jk}_{\m\al}p^\al_k]. \label{N20'}
\eea
It is readily observed that, with respect to the coordinates $\cS^i_\la$,
$\cF^i_\la$ (\ref{b4122}) and $\cR^\la_i$, $\cP^\la_i$ (\ref{N20'}), the
Lagrangian (\ref{N12}) reads 
\beq
\cL=\frac12 a^{\la\m}_{ij}\cF^i_\la\cF^j_\m +c', \label{cmp31}
\eeq
while the Lagrangian constraint space is given by the reducible constraints
\beq
\cR^\la_i= p^\la_i -
a^{\la\m}_{ij}\si_0{}^{jk}_{\m\al}p^\al_k=0. \label{zzz}
\eeq

Note that, in gauge theory on principal bundles, we have the canonical
splitting (\ref{N18}) where
$2\cF$ is the strength tensor \cite{book,sard94}.  The
Yang--Mills Lagrangian of this gauge theory is exactly of the form
(\ref{cmp31}) where $c'=0$. The Lagrangian of Proca fields is also of the form
(\ref{cmp31}) where
$c'$ is the mass term. This is an example of a  degenerate
Lagrangian system without gauge symmetries.

Given the linear map $\si$ (\ref{N17}) and a connection $\G$
(\ref{N16}), let us consider the Hamiltonian form
\ben
&& H= p^\la_idy^i\w\om_\la - [\G^i_\la
p^\la_i +\frac12 \si_0{}^{ij}_{\la\m}p^\la_ip^\m_j
+\si_1{}^{ij}_{\la\m}p^\la_ip^\m_j -c']\om=
\label{N22}\\
&& \quad (\cR^\la_i+\cP^\la_i)dy^i\w\om_\la - [(\cR^\la_i+\cP^\la_i)\G^i_\la
+\frac12
\si_0{}^{ij}_{\la\m}\cP^\la_i\cP^\m_j
+\si_1{}^{ij}_{\la\m}p^\la_ip^\m_j -c']\om.\nonumber
\een
One can show that the Hamiltonian forms (\ref{N22})
parametrised by connections $\G$ (\ref{N16})  constitute a complete set for
the Lagrangian  (\ref{N12}) \cite{book}.
It is readily observed that, if $\si_1$ is non-degenerate, so are the
Hamiltonian forms (\ref{N22}). 
Thus, for different $\si_1$, we have different complete sets of Hamiltonian
forms (\ref{N22}). Hamiltonian forms $H$
(\ref{N22}) of such a complete set differ from each other in the term
$\f^i_\la\cR^\la_i$, where $\f$ are the soldering forms (\ref{cmp21}). It
follows from the splitting (\ref{N20}) that this term vanishes on the
Lagrangian constraint space. 

\section{Geometry of antighosts}

We aim to obtain the Koszul--Tate resolution for the constraints
(\ref{zzz}).  Since these constraints are reducible, we need 
an infinite number of antighost fields in general \cite{fisch,kimura} (we
follow the terminology of \cite{kimura}). They are graded by the antighost
number
$r$ and the Grassmann parity $r\,{\rm mod}2$. Therefore, we should generalize
the notion of a graded manifold \cite{bart} to commutative graded
algebras generated  both by odd and even elements. 

Let $E=E_0\oplus E_1\to Z$ be the Whitney sum of vector bundles $E_0\to
Z$ and $E_1\to Z$ over a manifold $Z$. One can think of $E$ as
being a bundle of vector superspaces with a typical fibre $V=V_0\oplus V_1$
where transition functions of $E_0$ and $E_1$ are independent. Let us consider
the exterior bundle   
\be
\w E^*=\op\bigoplus^\infty_{k=0} (\op\w^k_Z E^*), 
\ee
which is the tensor bundle $\ot E^*$ modulo  elements 
\be
e_0e'_0 - e'_0e_0, \quad e_1e'_1 + e'_1e_1, \quad e_0e_1 - e_1e_0\quad
e_0,e'_0\in E_{0z}^*,
\quad e_1,e'_1\in E_{1z}^*, \quad z\in Z.
\ee 
Global sections of $\w E^*$ constitute a commutative graded algebra
$\cA(Z)$ modelled on  the locally free
$C^\infty(Z)$-module $E_0^*(Z)\oplus E_1^*(Z)$ of global sections of $E^*$.
This is the product of the commutative algebra $\cA_0(Z)$ of global sections of
the symmetric bundle $\vee E_0^*\to Z$ and the graded algebra $\cA_1(Z)$ of
global sections of the exterior bundle $\w E_1^*\to Z$. 
The pair
$(Z,\cA_1(Z))$ is a graded manifold \cite{bart}.  For the sake of brevity, we
agree to call 
$(Z,\cA(Z))$ a graded manifold, though its generating set contain an
even subset $\cA_0$. Accordingly, elements of $A(Z)$ are called graded
functions. Let us introduce the differential calculus in these functions.

We start from the $\cA(Z)$-module $\der \cA(Z)$ of graded derivations of
$\cA(Z)$. Recall that by a graded derivation of the commutative graded algebra
$\cA(Z)$ is meant an endomorphism of $\cA(Z)$ such that
\beq
 u(ff')=u(f)f'+(-1)^{\nw u\nw f}fu (f') \label{mm81}
\eeq
for the homogeneous elements $u\in \der\cA(Z)$ and $f,f'\in \cA(Z)$.
We use the notation $\nw
.$ for the Grassmann parity. 

We aim to show that graded derivations (\ref{mm81}) are represented by
sections of a vector bundle.
Let $\{c^a\}$ be the holonomic bases for $E^*\to Z$ with respect to some bundle
atlas $(z^A,v^i)$ of $E\to Z$ with transition functions $\{\rho^a_b\}$, i.e.,
$c'^a=\rho^a_b(z)c^b$. Then graded functions read
\beq
f=\op\sum_{k=0} \frac1{k!}f_{a_1\ldots
a_k}c^{a_1}\cdots c^{a_k}, \label{z785}
\eeq
where $f_{a_1\cdots
a_k}$ are local functions on $Z$, and we omit the symbol of an exterior product
of elements $c$. The coordinate transformation law of graded functions
(\ref{z785}) is obvious. 
Due to the canonical splitting
$VE= E\times E$, the vertical tangent bundle 
$VE\to E$ can be provided with the fibre bases $\{\dr_a\}$ dual of $\{c^a\}$.
These are fibre bases for $\pr_2VE=E$. Then
any derivation $u$ of $\cA(U)$ on a trivialization domain $U$ of $E$ reads
\beq
u= u^A\dr_A + u^a\dr_a, \label{mm83}
\eeq
where $u^A, u^a$ are local graded functions and $u$ acts on $f\in \cA(U)$ by
the rule
\be
u(f_{a_1\cdots
a_k}c^{a_1}\cdots c^{a_k})=u^A\dr_A(f_{a_1\cdots
a_k})c^{a_1}\cdots c^{a_k} +u^a
f_{a\ldots b}\dr_a\rfloor (c^a\cdots c^b). 
\ee
This rule implies the corresponding
coordinate transformation law 
\beq
u'^A =u^A, \qquad u'^a=\rho^a_ju^j +u^A\dr_A(\rho^a_j)c^j \label{lmp2}
\eeq
of derivations (\ref{mm83}).
Let us consider 
the vector bundle
$\cV_E\to Z$ which is locally isomorphic to the vector bundle
\be
\cV_E\mid_U\approx\w E^*\op\ot_Z(\pr_2VE\op\oplus_Z TZ)\mid_U,
\ee
and has the transition functions
\be
&& z'^A_{i_1\ldots i_k}=\rho^{-1}{}_{i_1}^{a_1}\cdots
\rho^{-1}{}_{i_k}^{a_k} z^A_{a_1\ldots a_k}, \\
&& v'^i_{j_1\ldots j_k}=\rho^{-1}{}_{j_1}^{b_1}\cdots
\rho^{-1}{}_{j_k}^{b_k}\left[\rho^i_jv^j_{b_1\ldots b_k}+ \frac{k!}{(k-1)!} 
z^A_{b_1\ldots b_{k-1}}\dr_A(\rho^i_{b_k})\right] 
\ee
of the bundle coordinates $(z^A_{a_1\ldots a_k},v^i_{b_1\ldots b_k})$,
$k=0,\ldots$. These transition functions
fulfill the cocycle relations. It is readily observed that, for any
trivialization domain $U$, the
$\cA$-module $\der\cA(U)$ with the transition functions (\ref{lmp2}) is
isomorphic to the $\cA$-module of local sections of $\cV_E\mid_U\to U$.
One can show that, if $U'\subset U$ are open
sets, there is the restriction morphism $\der\cA(U)\to
\der\cA(U')$. It follows that, restricted to an open subset $U$, every
derivation $u$ of
$\cA(Z)$ coincides with some local section $u_U$ of $\cV_E\mid_U\to U$, whose
collection $\{u_U, U\subset Z\}$ defines uniquely a global section of
$\cV_E\to Z$, called a graded vector field on $Z$. 

The $\w E^*$-dual $\cV^*_E$ of $\cV_E$ is a vector bundle over $Z$
whose sections 
constitute the $\cA(Z)$-module of  exterior graded
1-forms $\f=\f_A dz^A + \f_adc^a$.
Then
the morphism $\f:u\to \cA(Z)$ can be seen as the interior product 
\beq
u\rfloor \f=u^A\f_A + (-1)^{\nw{\f_a}}u^a\f_a. \label{cmp65}
\eeq
Graded $k$-forms $\f$ are defined as sections
of the graded exterior bundle $\w^k_Z\cV^*_E$ such that
\be
 \f\w\si =(-1)^{\nm\f\nm\si +\nw\f\nw\si}\si \w \f,  
\ee
where $|.|$ is the form degree.
The interior product (\ref{cmp65})
is extended to higher graded forms by the rule  
\be
u\rfloor (\f\w\si)=(u\rfloor \f)\w \si
+(-1)^{\nm\f+\nw\f\nw{u}}\f\w(u\rfloor\si). 
\ee
The graded exterior differential
$d$ of graded functions is introduced by the condition 
$u\rfloor df=u(f)$
for an arbitrary graded vector field $u$, and  is
extended uniquely to higher graded forms by the rules
\be
d(\f\w\si)= (d\f)\w\si +(-1)^{\nm\f}\f\w(d\si), \qquad  d\circ d=0.
\ee

\section{The Koszul--Tate resolution}

Let us turn to the splitting (\ref{N20}) and introduce the projection
operators
\be
P^{\la k}_{i \nu}= a_{ij}^{\la\m}\si_0{}^{jk}_{\m\nu}, 
\qquad  R^{\la k}_{i \nu}=(\dl_i^k\dl^\la_\nu -
a_{ij}^{\la\m}\si_0{}^{jk}_{\m\nu})
\ee
such that 
\beq
P^{\la k}_{i \nu}\cR^\nu_k=0, \qquad P^{\la k}_{i \nu}\cR^\nu_k=\cR^\la_i.
\label{xxx}
\eeq

To construct the vector bundle $E$ of antighosts, let
us consider the vertical tangent bundle $V_Y\Pi$ of $\Pi\to Y$.
Let us chose the bundle $E$ as the Whitney sum of the
bundles $E_0\oplus E_1$ over $\Pi$ which are the infinite Whitney sum
over
$\Pi$ of the copies of
$V_Y\Pi$.  We have
\be
E= V_Y\Pi\op\oplus_\Pi V_Y\Pi\op\oplus_\Pi\cdots.
\ee
This bundle is provided with the holonomic coordinates $(t,y^i,p_i^\la,\dot
p_i^{\la(r)})$, $r=0,1,\ldots$, where  $(t,y^i,p_i^\la,\dot
p_i^{\la(2l)})$ are coordinates on $E_0$, while 
$(t,y^i,p_i^\la,\dot p_i^{\la(2l+1)})$ are those on $E_1$. By
$r$ is meant the antighost number.  The
dual of
$E\to V^*Q$ is
\be
E^*= V^*_Y\Pi\op\oplus_\Pi V^*_Y\Pi\op\oplus_\Pi\cdots.
\ee
It is  endowed with the
associated fibre bases
$\{c_i^{\la(r)}\}$, $r=1,2,\ldots$, such that
$c_i^{\la(r)}$ have the same linear coordinate
transformation law as the coordinates $p_i^\la$. The corresponding
graded vector fields and graded forms are introduced on $\Pi$ as sections of
the vector bundles $\cV_E$ and $\cV^*_E$, respectively. 

The $C^\infty(\Pi)$-module $\cA(\Pi)$ of graded functions is graded by the
antighost number as 
\be
\cA(\Pi)=\op\oplus_{r=0}^\infty \cN^r, \qquad \cN^0=C^\infty(\Pi).
\ee
Its terms $\cN^r$ constitute a complex
\beq
0\leftarrow C^\infty(\Pi) \leftarrow \cN^1 \leftarrow \cdots \label{mm90}
\eeq
with respect to the Koszul--Tate  differential
\ben
&& \dl: C^\infty(V^*Q)\to 0, \nonumber \\
&& \dl(c^{\la(2l)}_i)=P^{\la k}_{i \nu}c^{\nu(2l-1)}_k,
\qquad l>0,
\label{mm91}\\ 
&& \dl(c^{\la(2l+1)}_i)=R^{\la k}_{i \nu} c^{\nu(2l)}_k,
\qquad l>0,
\nonumber\\ &&  \dl(c^{\la(1)}_i)=R^{\la k}_{i \nu} p_k^\nu.
\nonumber
\een
The nilpotency property $\dl\circ\dl=0$ of this differential is the corollary
of the relations (\ref{xxx}).

It is readily observed that the complex (\ref{mm90}) with respect to the
differential (\ref{mm91}) has the homology
groups 
\be
H_{k>1}=0, \qquad H_0=C^\infty(V^*Q)/I_{N_L}=C^\infty(N_L), 
\ee
where $I_{N_L}$ is an ideal of smooth functions on $\Pi$ which vanishes on
the Lagrangian constraint space $H_L$. Thus, this is a desired Koszul--Tate
resolution of the constraints (\ref{zzz}) defined by the Lagrangian
(\ref{N12}).

Note that, in different particular cases of the degenerate quadratic Lagrangian
(\ref{N12}), the complex (\ref{mm90}) may have a subcomplex, which is also
the Koszul--Tate resolution. For instance, if the fibre metric $a$ in $VQ\to
Q$ is diagonal with respect to a holonomic atlas of $VQ$, the constraints
(\ref{zzz}) are irreducible and the complex (\ref{mm90}) contains a
subcomplex which consists only of the antighosts
$c_i^{\la(1)}$.

\end{document}